\documentstyle[12pt]{article}
\begin{document}
\begin{titlepage}
\rightline {Si-94-11 \  \  \  \   }

\vskip 2truecm
\centerline{\Large\bf
Local Differential Geometry}
\centerline{\Large \bf as a Representation of the SUSY  Oscillator}

\vskip 1.0truecm

\centerline{PACS Nos.: 11.30, 3.65, 2.40}

\vskip 2.0truecm

\centerline{\bf H.-P. Thienel}

\vskip 0.7truecm

\centerline{Universit\"at-GH Siegen, Fachbereich 7, D-57068 Siegen, Germany}
\centerline{e-mail: thienel@sicip1.physik.uni-siegen.de}

\vskip 2.0truecm

\abstract

\noindent The choice of a coordinate
chart on an analytical $R^n$ ($R^n_{\rm a}$) provides
a representation of the $n$-dimensional SUSY oscillator.
The 1-parameter group of dilations provides a Euclidean evolution
moving the system through a sequence of charts, that at
each instant supplies a Hilbert space by Cartan's exterior algebra endowed with
a suitable scalar product.
Stationary states and coherent states are eigenstates of
the Lie derivatives generating the dilations and the translations
respectively.

\end{titlepage}

\section*{I. Introduction}

In this paper the presence of differential geometric structures
in a special but important quantum system will be worked out.
The $n$-dimensional SUSY oscillator [1-6]
turns out to be a perfect manifestation of the
differential geometry of
the most simple manifold: analytical $R^n$ ($R^n_{\rm a}$),
where a coordinate system exists globally.
The  quantum structures are due to a selected chart on this manifold.
Hence, it is important to observe
 that coordinates are not auxiliary quantities.
In this respect
this representation is similar to
the (anti-)holomorphic representations of Fock \cite{foc}, Bargmann
\cite{bar} and Berezin \cite{be1}.
However, as a major difference,
the present work involves exclusively real quantities.

The paper is divided into two parts.
We start by restating basic differential
geometrical notions \cite{cho} and show
that the choice of a coordinate system on
$R^n_{\rm a}$ amounts to finding  familiar geometrical objects that are
  reinterpreted as operators and state vectors from a Fock space point of view.
In particular, forms consisting of commuting coordinates and anticommuting
differentials are suited for a representation of a bosonic/fermionic Fock
space. In contrast to \cite{wit1}, where square integrable
differential forms are employed,
and  the metric
provides the scalar product,
the treatment of bosonic and fermionic degrees of freedom is perfectly
analogous in our approach. This requires, as the only
additional ingredient from
a geometrical point of view, the introduction of a scalar product for the
forms. A metric is not needed on  $R^n_{\rm a}$.
{}From the discussion of the exterior
derivative  the algebra of the SUSY oscillator  is derived. The
corresponding Hamiltonian is, on the one hand,  the Lie derivative generating
the dilations on $R^n_{\rm a}$, on the other hand, it is
a total derivative giving rise to a euclidean Schr\"odinger equation.
The euclidean evolution enables
the formalism to be maintained strictly real.
Stationary states are defined to be the eigenstates of the Hamiltonian.
Analogously, coherent states are defined as eigenstates
of the Lie derivatives generating the translations on $R^n_{\rm a}$.

For an illustration, in the second part,
we concentrate on $R^1_{\rm a}$, i.e. the 1-dimensional SUSY oscillator,
which is among a number of
physical  SUSY quantum mechanical systems
\cite{gen},\cite{lan},\cite{nieto1},\cite{nieto2}.
We begin  with a discussion of stationary states.
Finally, coherent states are found to exhibit classical  motion within
the inverted oscillator potential
propagating in euclidean time.

\section*{II. Geometry of $R^n_{\rm a}$}
\section*{ and  the $n$-dimensional SUSY oscillator}

Consider $R^n$ as a differentiable manifold \cite{cho}.
We pick an arbitrary point $\in R^n$ and take it as the origin of
coordinates
$\{ x^i\} $ that parametrize $R^n$ entirely. We restrict
ourselves to functions  that are analytical around the origin
\begin{equation}
f(\{ x^i\})=f_0+f_i x^i+ f_{ij}x^i x^j+..., \qquad f_{ij...}={\rm const.} \in R
\label{1}
\end{equation}
defining $R^n_{\rm a}$.
Immediately, we have two kinds of operators acting on these
functions (from the left).
The coordinate function $x^i$ is an operator that acts
on $f(\{ x^i\})$
by multiplication. $\partial_{x^i}:=\partial/\partial x^i$,
generating
displacements on $R^n_{\rm a}$, acts on $f(\{ x^i\})$ by performing the partial
derivative with
respect to $x^i$. They yield the algebra
\begin{equation}
[x^i,x^j]_-=[\partial_{x^i},\partial_{x^j}]_-=0,
\qquad [\partial_{x^i},x^j]_-=\delta^j_i,
\label{2}
\end{equation}
due to
$  \partial_{x^k} x^j=\delta_k^j$.
These and all following commutators are graded by the form degrees of the
entries,
i.e.  $[A,B]=[A,B]_+$,  if both $A$ and $B$ have odd form degree and
$[A,B]=[A,B]_-$ otherwise.
Eq.(\ref{2}) is regarded as a bosonic algebra of creators
$x^i$ and annihilators $\partial_{x^i}$. From this point of view the function
in eq.(\ref{1}) represents an element of a bosonic
Fock space, if we interprete the
Taylor series as a power expansion in the
creators $x^i$ applied to a constant number.
Having defined analytical $0$-forms in eq.(\ref{1}),
we consider now analytical  $p$-forms on $R^n_{\rm a}$.
In our
chart they can be expressed as
\begin{equation}
F^{(p)}=F_{i_1,...,i_p}(\{ x^i\})dx^{i_1}....dx^{i_p}\in \Lambda^p R^n_{\rm a},
\label{3}
 \end{equation}
with the factors $F_{i_1,...,i_p}(\{ x^i\})$
being analytical functions as in eq.(\ref{1}).
The product of the $dx^i$ (to be read $(dx^i)$ in the present terminology)  is
anticommutative, often symbolized by a wedge that is omitted here.
Just as a function is a series in  $x^i$,  a  form is defined
as  a series in the Grassmann numbers
$dx^i$, which is finite in $dx^i$
for finite dimension $n$ in contrast to eq.(\ref{1}),
\begin{eqnarray}
 \Psi (\{ x^i\},\{ dx^i\})&=&F_0(\{ x^i\})+F_j(\{ x^i\}) dx^j
+ F_{jk}(\{ x^i\})dx^j dx^k+...  \nonumber
\\ \quad \in \Lambda R^n_{\rm a}
&=&\bigoplus_{p=0}^n \Lambda^p R^n_{\rm a},
\label{6}
\end{eqnarray}
spanning Cartan's exterior algebra, which is
a combined Fock space
with $n$ bosonic
$x^i$ and $n$ fermionic $dx^i$.
The contraction of a $p$-form,
with a vector $v=v^i \partial_{x^i}\in TR^n_{\rm a}$
yields a  ($p-1$)-form.
This is the interior product operation usually indicated by $i_v$.
We write alternatively
\begin{equation}
v^i \partial_{dx^i}:=v^i \frac{\partial}{\partial dx^i}\equiv i_v ,
\label{4}
\end{equation}
where $\partial_{dx^i}$ is the left derivative  with
respect to $dx^i$ \cite{be1} carrying form degree $-1$.
The duality of $\partial_{x^i} \in TR^n_{\rm a}$ and
$dx^j \in T^*R^n_{\rm a}$
is expressed by
$ \partial_{dx^k} dx^j=\delta_k^j$ giving rise to a new set of
algebraic relations
\begin{equation}
[dx^i,dx^j]_+=[\partial_{dx^i},\partial_{dx^j}]_+=0,
\qquad [\partial_{dx^i},dx^j]_+=\delta^j_i,
\label{5}
\end{equation}
 Obviously, we obtained a fermionic  algebra analogous to eq.(\ref{2})
consisting of  creators $dx^i$  and annihilators $\partial_{dx^i}$. Any form
can be created
by applying  1-forms $dx^i$ (from the left)
and  bosonic factors
$ x^i  $
 to a constant number.
We complete eqns. (\ref{2}) and (\ref{5}) by the
the mixed commutators
\begin{equation}
[x^i,dx^j]_-=[x^i,\partial_{dx^j}]_-=[\partial_{x^i},dx^j]_-=
[\partial_{x^i},\partial_{dx^j}]_-=0.
\label{7}
\end{equation}
We now define the adjoints of the  fundamental  operators
\begin{equation}
 (x^i)^+:=\partial_{x^i},\qquad
 (dx^i)^+:=\partial_{dx^i},
\label{8}
\end{equation}
while real numbers are  self-adjoint.
The adjoint of an arbitrary operator is obtained
by decomposing it into the fundamental
operators above and applying the usual rules
$ (A+B)^+=A^+ +B^+$, $ (AB)^+=B^+A^+$ and $(A^+)^+=A$.
In particular, the adjoint operation
applied to the Fock space vector eq.(\ref{6})
supplies the dual vector $\Psi^+(\{ \partial_{dx^i}\} ,\{ \partial_{x^i}\})$
from the dual of $\Lambda R^n_{\rm a}$.
We denote the scalar product of two forms $\Psi$, $\Xi \in \Lambda R^n_{\rm a}$
by
\begin{equation}
\langle \Psi,\Xi\rangle :=(\Psi^+\Xi)|_{x^i=dx^i=0}.
\label{12}
\end{equation}
It is calculated by performing
all the derivations of $\Psi^+$ on $\Xi$
and putting the remaining factors $x^i$ and $dx^i$
to zero.
This scalar product is symmetric and induces a positive definite norm.
Therefore, endowed with this scalar product the Fock
space $\Lambda R^n_{\rm a}$
becomes a Hilbert space. In \cite{dub}   a similar scalar
product was introduced for the space of functionals in quantum field theory.

Until now, we implicitly used the exterior derivative $d$ that transforms a
bosonic $x^i$ into a fermionic $dx^i$.
It can be written as
$d=dx^i\partial_{x^i}$
and is obviously nilpotent
$ [d,d]_+=0.$
We can construct its adjoint which turns a fermionic $dx^i$ into a bosonic
$x^i$, hence
$d^+=x^i\partial_{dx^i}$
being also nilpotent
$ [d^+,d^+]_+=0.$
This operator is also defined in \cite{bra}
and slightly different in
\cite{cho}
for the proof of Poincar\'e's lemma.
We calculate the commutator of $d$ and $d^+$, defining the self-adjoint
Hamiltonian
\begin{equation}
 H:=[d,d^+]_+=x^i \partial_{x^i}+dx^i\partial_{dx^i}
\label{13}
\end{equation}
The first term counts the powers in the coordinates of an expression that
it is applied on. So we define the boson number operator
$N:=x^i\partial_{x^i}$,
which is  bosonic.
The second term
counts the form degree, if it is applied to a $p$-form.
Accordingly, we define the fermion number operator
$P:=dx^i\partial_{dx^i}$,
which is also bosonic.
$d$ and $d^+$ are conserved
\begin{equation}
[d,H]_-=[d^+,H]_-=0.
\label{14}
\end{equation}
Equations (\ref{13}) and (\ref{14}) represent the algebra
of an $n$-dimensional SUSY oscillator with the supersymmetry
charges $d$ and $d^+$ \cite{nic,wit}. The Hamiltonian reads
$H=(N+n/2)+(P-n/2),$   the zero-point energies reinserted.

Generally, a Lie derivative with respect to the vector field $v=v^i
\partial_{x^i}$ can be written as
${\cal L}_v=[d, v^i \partial_{dx^i}]_+$.
Therefore, the Hamiltonian $H$
is a Lie derivative with respect to the vector field
$x^i \partial_{x^i}$ with integral lines which are rays
emanating from the origin.
In fact, $H\equiv {\cal L}_{x^i \partial_{x^i}}$
is a total derivative with respect to a
parameter $\tau$ \cite{cho}
\begin{equation}
-{d \over d\tau} \Psi(\tau)=H\Psi(\tau)
\label{15}
\end{equation}
for any
\begin{equation}
\Psi(\{ x^i(\tau)\} ,\{ dx^j(\tau)\})
=\Psi(\{ x^i e^{-\tau}\} ,\{ dx^j e^{-\tau} \} )
=e^{-\tau H}\Psi(\{ x^i \} ,\{ dx^j \} )
\in \Lambda R^n_{\rm a},
\label{16}
\end{equation}
where $x^i \equiv x^i(0)$.
$\tau$ parametrizes a sequence of charts on $R^n_{\rm a}$ each representing
the quantum system at an instant $\tau$.
The rule for the adjoint operation is to be applied within each chart, hence
\begin{equation}
 (x^i(\tau))^+=\partial_{x^i(\tau)},\qquad
 (dx^i(\tau))^+=\partial_{dx^i(\tau)}.
\label{88}
\end{equation}
As a consequence, the norm is independent of $\tau < \infty$.
The compatibility of eqns.(\ref{8}) and (\ref{88})
implies $\tau^+=-\tau$, therefore, while in the geometric
framework $\tau$ is to be regarded a real number , in the sense of the
 scalar product, relating the Fock space and its dual, it is not.
Thus, unlike in quantum mechanics,
the sequence of Hilbert spaces, representing the evolution of the
system, is not a Hilbert space itself.

Observe, that for $\tau = \infty$ all states but the constant functions,
representing the only non-vanishing  cohomology-class on $R^n_{\rm a}$,
vanish.

Obviously, we have found a euclidean
Schr\"odinger type equation determining the $\tau$-propagation of any $\Psi$
similar to quantum mechanics. (In fact, by
performing
an analytical continuation  $\tau \to it$, using
complex coordinates and constants,
where the adjoint of a constant is its complex conjugate,
one ends up with the real-time quantum
mechanics governing the $n$-dimensional SUSY oscillator in creator/annihilator
language.)

We define  stationary states to  have the evolution
$\phi(\tau)=\phi e^{-\epsilon\tau}$ and therefore to be eigenstates of $H$
\begin{equation}
H\phi=\epsilon \phi.
\label{18}
\end{equation}

 The definition for  coherent states
here is inspired by that of the stationary states.
Consider the Lie derivative with respect to a constant vector field
\begin{equation}
{\cal L}_{c^i \partial_{x^i}} =[d,c^i\partial_{dx^i}]_+=c^i \partial_{x^i},
\label{19}
\end{equation}
which turns out to be a simple directional derivative on any element of
Cartan's
exterior algebra. The coherent states are defined to be solutions of the
corresponding eigenvalue problem
\begin{equation}
c^i \partial_{x^i}\kappa =\gamma  \kappa ,\qquad
\gamma \in R.
\label{20}
\end{equation}

Since $[{\cal L}_v,d]_-=0$, from a given  eigenstate of ${\cal L}_v$
another one is immediately found
by application of $d$, such that eigenstates of ${\cal L}_v$ will always come
in SUSY pairs.

We conclude this section
by remarking,
that this representation is the one where the creation operator $x^i$  is
diagonal. Its continuous
spectrum is given by the points on $R^n$. The corresponding
eigenstates are not normalizable as is well-known \cite{mil} and easily shown.
This is analogous
to quantum mechanics where the position and momentum operators
provide representations without having  square-integrable eigenstates
themselves.

\section*{III. Stationary states and coherent states on $R^1_{\rm a}$}

The  eigenvalue problem of  the Hamiltonian on $R^1_{\rm a}$ is
\begin{equation}
H\phi_{\epsilon}=\epsilon\phi_{\epsilon},
\qquad H=x\partial_{x} +dx\partial_{dx}
\label{23}
\end{equation}
Due to $[H,P]_-=0$, there are eigenstates of definite form degree,
indicated by superscripts in parentheses.
We  begin by  choosing form degree zero, such that
\begin{equation}
x\partial_{x}\phi^{(0)}_{\epsilon}=\epsilon \phi^{(0)}_{\epsilon}.
\label{23a}
\end{equation}
Obviously,  $\phi^{(0)}_{\epsilon} \sim x^{\epsilon}$, where
a priori  $\epsilon$ is any real number. However, the condition that
all states have to be analytic, immediately rules out all solutions of
(\ref{23a}) with the exception of
$\phi^{(0)}_{\epsilon}=(1/ \sqrt{(\epsilon )!})x^{\epsilon }$,
($\epsilon =0,1,2,...$).
 There is however a
degeneracy of $H$. Application of $d$ to eq.(\ref{23}) yields
\begin{equation}
Hd\phi^{(0)}_{\epsilon}=\epsilon d\phi^{(0)}_{\epsilon},
\label{26}
\end{equation}
such that  $d\phi^{(0)}_{\epsilon}$
is as well an eigenfunction of $H$ corresponding to the same eigenvalue as
$\phi^{(0)}_{\epsilon}$. Therefore, by application of $d$ to the
set of eigenstates $\phi^{(0)}_{\epsilon}$,
we get
a second series of eigenstates
$\phi^{(1)}_{\epsilon}=(1/ \sqrt{(\epsilon -1)!})x^{\epsilon -1}dx$,
($ \epsilon =1,2,...$).
The only state without degeneracy is the vacuum $\phi^{(0)}_0=1$ satisfying
$d 1=d^+ 1=0,$
implying that supersymmetry is not spontaneously broken.
The complete set of
eigenstates  consists of both $\{ \phi^{(0)}_{\epsilon} \} $ and
$\{ \phi^{(1)}_{\epsilon} \} $, obeying the orthonormality relation
\begin{equation}
\langle \phi^{(p)}_{\epsilon}, \phi_{\epsilon'}^{(p')}\rangle =
\delta_{\epsilon \epsilon'}\delta_{p p'} \qquad p=0,1.
\label{27}
\end{equation}

The eigenvalue problem for coherent states on $R^1_{\rm a}$ reads
\begin{equation}
\partial_{x}\kappa_{\gamma} =\gamma  \kappa_{\gamma}, \qquad \gamma \in R.
\label{29}
\end{equation}
Note that  $[\partial_{x},P]_-=0$,
such
that again we can find solutions of definite form degree $p$, coming
in pairs of supersymmetric partners due to $[\partial_x,d]_-=0$
\begin{equation}
\kappa_{\gamma}^{(p)}=e^{-{1 \over 2} \gamma^2 }
e^{\gamma x} (dx)^p
\label{32}
\end{equation}
after normalization.
Coherent states of different
form degree are clearly orthogonal. But the scalar
product of two arbitrary $p$-form coherent states is
\begin{equation}
\langle \kappa^{(p)}_{\gamma},\kappa^{(p')}_{\gamma'}\rangle
=e^{-{1 \over 2}(\gamma -\gamma')^2} \delta_{p p' }.
\label{33}
\end{equation}

For general coherent states
there exist three major definitions in the literature,
which coincide in the case of the harmonic oscillator:
(i) the displacement operator definition,
(ii) the annihilation operator eigenstate definition,
(iii) the minimum uncertainty definition \cite{nieto3}.
By employing supergroups, also coherent states in  SUSY quantum mechanical
systems can meet all of these definitions \cite{fat}.
The
coherent states in the present scenario
deviate from these ''super-coherent states'', since Grassmann eigenvalues
are not involved.
The definition of coherent states eq.(\ref{20})  corresponds
to (ii), but
in the following, we will show that they are also minimum-uncertainty states
(iii).

Consider the operators
$q=(1/\sqrt{2}) (x+\partial_{x})=q^+$
and $\pi=(1/\sqrt{2}) (x-\partial_{x})=-\pi^+$ with
$[q,\pi]_-=-1$.
The $\tau$ evolution of a coherent state of definite form degree yields
\begin{equation}
\kappa_{\gamma}^{(p)}(\tau)=e^{-{1 \over 2} \gamma^2 }
e^{\gamma e^{-\tau}x} (dx e^{-\tau})^p.
\label{44}
\end{equation}
The expectation values
\begin{equation}
\langle \kappa_{\gamma}^{(p)}(\tau),q\kappa_{\gamma}^{(p)}(\tau)\rangle =
\sqrt{2} \gamma \cosh\tau,
\qquad \langle \kappa_{\gamma}^{(p)}(\tau),
\pi\kappa_{\gamma}^{(p)}(\tau)\rangle =\sqrt{2} \gamma \sinh\tau
\label{45}
\end{equation}
satisfy the classical equations of motion in $\tau$
within an inverted oscillator potential with energy $>0$, if $q$
corresponds to the position and $\pi$ to the momentum in the classical phase
space.
By analytical continuation of $\tau \to it $ and $\pi \to -i $  times momentum,
we recover the classical motion of the harmonic oscillator.

Finally, $\Delta q=\Delta \pi =1/\sqrt{2}$ for coherent states,
with $(\Delta q)^2= \langle\Psi,( q^2-\langle\Psi, q \Psi\rangle^2)\Psi\rangle$
and
$(\Delta \pi)^2=\langle\Psi, (\langle\Psi,\pi \Psi\rangle^2-\pi^2)\Psi\rangle$,
implying
$\Delta q\Delta \pi  =  1/2$ for any $\tau$.

\section*{IV. Conclusions}

Since the geometrically most simple situation of an $R^n_{\rm a}$
 already supplies a
fundamental physical
application by the above formalism, it is tempting to consider
extensions of the geometry in view of extending the range
of physical applicability.
In particular,
 the promotion of the presented concepts to non-trivial
manifolds appears to be attractive having in mind
a correspondence between cohomology classes and SUSY vacua.

\vskip 1.0truecm
{\bf Acknowledgement:}

\noindent I am indebted to H.-D. Dahmen,
 S. Marculescu and M. Reuter for important hints.

\end{document}